\RequirePackage[]{graphicx} 
\graphicspath{{./Figures/}}

\documentclass[]{arxsub}

\usepackage[utf8]{inputenc}
\usepackage{graphicx}
\usepackage{environ}
\usepackage{listings}
\usepackage{booktabs}
\usepackage{blindtext}
\usepackage{url}

\title{Assessment of Deep Learning Segmentation for Real-Time Free-Breathing Cardiac Magnetic Resonance Imaging \newline at Rest and Under Exercise Stress}
\shorttitle{Assessment of Deep Learning Segmentation for Real-Time CMR}
\corres{Martin Uecker, Graz University of Technology,
Institute of Biomedical Imaging, Stremayrgasse~16/3, 8010~Graz, AUSTRIA, \email{uecker@tugraz.at}}

\author[1$\dagger$]{Martin Schilling}{}
\equalcont{These authors contributed equally to this work.}

\author[1,2,3$\dagger$]{Christina Unterberg-Buchwald}{}
\equalcont{These authors contributed equally to this work.}

\author[1]{Joachim Lotz}{}
\author[1,2,4]{Martin Uecker}{}

\jnlcitation{\cname{%
\author{M. Schilling},
\author{C. Unterberg-Buchwald},
\author{J. Lotz}, and
\author{M. Uecker}} (\cyear{2023}),
\ctitle{Assessment of Deep Learning Segmentation for Real-Time Free-Breathing Cardiac Magnetic Resonance Imaging}, \cjournal{}, \cvol{}.
}

\authormark{M. Schilling \textsc{et al.}}
\date{}

\address[1]{Institute for Diagnostic and Interventional Radiology, University Medical Center Göttingen, Göttingen, Germany}
\address[2]{German Centre for Cardiovascular Research (DZHK), Partner Site Göttingen, Göttingen, Germany}
\address[3]{Clinic of Cardiology and Pneumology, University Medical Center Göttingen, Göttingen, Germany}
\address[4]{Institute of Biomedical Imaging, Graz University of Technology, Graz, Austria}

\articletype{Research Article}

\finfo{}

\abstract{
\section{Purpose}
In recent years, a variety of deep learning networks for cardiac MRI (CMR) segmentation have been developed and analyzed.
However, nearly all of them are focused on cine CMR under breathold.
In this work, accuracy of deep learning methods is assessed for volumetric analysis (via segmentation) of the left ventricle
in real-time free-breathing CMR at rest and under exercise stress.
\section{Methods}
Data from healthy volunteers (n=15) for cine
and real-time free-breathing CMR at rest and under exercise stress
were analyzed retrospectively.
Exercise stress was performed using an ergometer in the supine position.
Segmentations of two deep learning methods,
a commercially available technique (comDL) and an openly available
network (nnU-Net), were compared to a reference model created via the
manual correction of segmentations obtained with comDL.
Segmentations of left ventricular endocardium (LV), left ventricular myocardium (MYO),
and right ventricle (RV) are compared for both end-systolic and end-diastolic phases
and analyzed with Dice's coefficient (DC).
The volumetric analysis includes the cardiac function parameters LV end-diastolic volume (EDV), LV end-systolic volume (ESV), and LV ejection fraction (EF),
evaluated with respect to both absolute and relative differences.
\section{Results}
For cine CMR, nnU-Net and comDL achieve Dice's coefficients above 0.95 for LV and 0.9 for MYO, and RV.
For real-time CMR, the accuracy of nnU-Net exceeds that of comDL overall.
For real-time CMR at rest, nnU-Net achieves Dice's coefficients of 0.94 for LV, 0.89 for MYO, and 0.90 for RV
and the mean absolute differences between nnU-Net and the reference
are 2.9\,mL for EDV, 3.5\,mL for ESV and 2.6\,\% for EF.
For real-time CMR under exercise stress, nnU-Net achieves Dice's coefficients of 0.92 for LV, 0.85 for MYO, and 0.83 for RV
and the mean absolute differences between nnU-Net and reference
are 11.4\,mL for EDV, 2.9\,mL for ESV and 3.6\,\% for EF.
\section{Conclusion}
Deep learning methods designed or trained for cine CMR segmentation
can perform well on real-time CMR.
For real-time free-breathing CMR at rest, the performance of deep learning methods
is comparable to inter-observer variability in cine CMR and is usable for fully automatic segmentation.
For real-time CMR under exercise stress, the performance of nnU-Net could promise a higher degree of automation in the future.
}

\begin{document}

\maketitle

\section{Introduction}

The fast and reliable evaluation of cardiac function is an essential part of cardiac MRI (CMR), significant
for patient diagnostics, disease analysis, therapy evaluation, follow-up,
and risk estimation \cite{White_Circ.J._1987,Norris_Eur.HeartJ._1992}.
The main quantitative parameters of cardiac function are the left ventricular blood volume (LV),
the volume of left ventricular myocardium (MYO), and the right ventricular blood volume (RV).
These parameters are usually calculated by acquiring and segmenting a stack of cross-sectional images in short-axis view.

Advances in image reconstruction have enabled the continuous acquisition of high-quality images
in real time, i.e. during free-breathing and independent of ECG-synchronization
\cite{Uecker_NMRBiomed._2010,Feng_Magn.Reson.Med._2013,Saybasili_Magn.Reson.Imaging_2014}.
In CMR, real-time MRI has emerged as a viable alternative for patients with arrhythmia \cite{Laubrock_Eur.J.Radiol.Open_2022},
for measurements using exercise stress \cite{Steinmetz_Circ.Cardiovasc.Imaging_2021,Li_Magn.Reson.Imaging_2021,Backhaus_Circ.J._2023},
as well as for real-time guidance in cardiac catheter interventions
\cite{Eitel_Curr.Cardiol.Rep._2014,Unterberg-Buchwald_J.Cardiov.Magn.Reson._2017,Campbell-Washburn_J.Magn.Reson.Imaging_2017,Franson_J.Imaging_2021}.

The delineation of the LV boundary is an important step for the determination of end-diastolic volume (EDV), end-systolic volume (ESV), and ejection fraction (EF).
However, manual segmentation of images is tedious and affected by inter- and intra-observer variability
\cite{Bernard_IEEETrans.Med.Imag._2018,Bai_J.Cardiov.Magn.Reson._2018,Bhuva_CircImaging_2019}.
In recent years, deep learning methods have been introduced into clinical practice for the generation of base contours, which are manually corrected as required.
In research, a variety of deep learning methods have been developed for segmentation in CMR \cite{Chen_Front.Cardiovasc.Med._2020}.
However, nearly all of these methods are focused on conventional breathold cine imaging \cite{Shoaib_ComputationalIntelligenceandNeuroscience_2023}.
In real-time MRI, automatic segmentation becomes more important, because
a series of heart beats is acquired instead of a single cine loop.
Real-time exercise stress studies pose an additional challenge for automatic segmentation due to the tendency
for inferior image quality \cite{Li_Magn.Reson.Imaging_2021,Morales_J.Cardiov.Magn.Reson._2023} depending on an increased heart rate and breathing motion.
Recent works covering real-time MRI \cite{Yang_BioMedRes.Int._2019,Qi_Physiol.Meas._2022} have used custom neural networks
trained specifically for the application on real-time CMR, though have faced the problem of limited training data availability.

This study aims to investigate the feasibility of using deep learning methods
trained on cine CMR data for the automatic segmentation of real-time
free-breathing CMR images. It will evaluate the performance of the
methods under the conditions of both rest and during exercise stress.

\section{Methods}
\subsection{Overview}

We analyzed cine and real-time measurements of
healthy volunteers (n=15) acquired at rest and under exercise stress
using a highly undersampled radial bSSFP sequence with NLINV reconstruction \cite{Uecker_Magn.Reson.Med._2010,Zhang_J.Cardiov.Magn.Reson._2010}
with a temporal resolution of 33\,ms resolution.

We evaluated segmentations obtained with two deep learning methods, (1)
the automatic contour detection designed for cine CMR (comDL) that
is included in the commercially available software Medis (version 4.0.56.4, QMass® 8.1, Medical Imaging Systems, Leiden, Netherlands)
and (2) nnU-Net \cite{Isensee_Nat.Methods_2021},
which was pre-trained on the cine CMR dataset of the Automated Cardiac Diagnosis Challenge (ACDC).
Segmentations were compared to a reference
model and standard clinical routine procedure
of manually correcting the results obtained with comDL.

The accuracy of segmentation was evaluated for images in end-diastolic and end-systolic phases.
The cardiac function parameters EDV, ESV, and EF were derived from neural network segmentation
and compared to reference values derived from the manually corrected contours.

\begin{figure*}[ht]
	\centering
	\includegraphics[width=\textwidth]{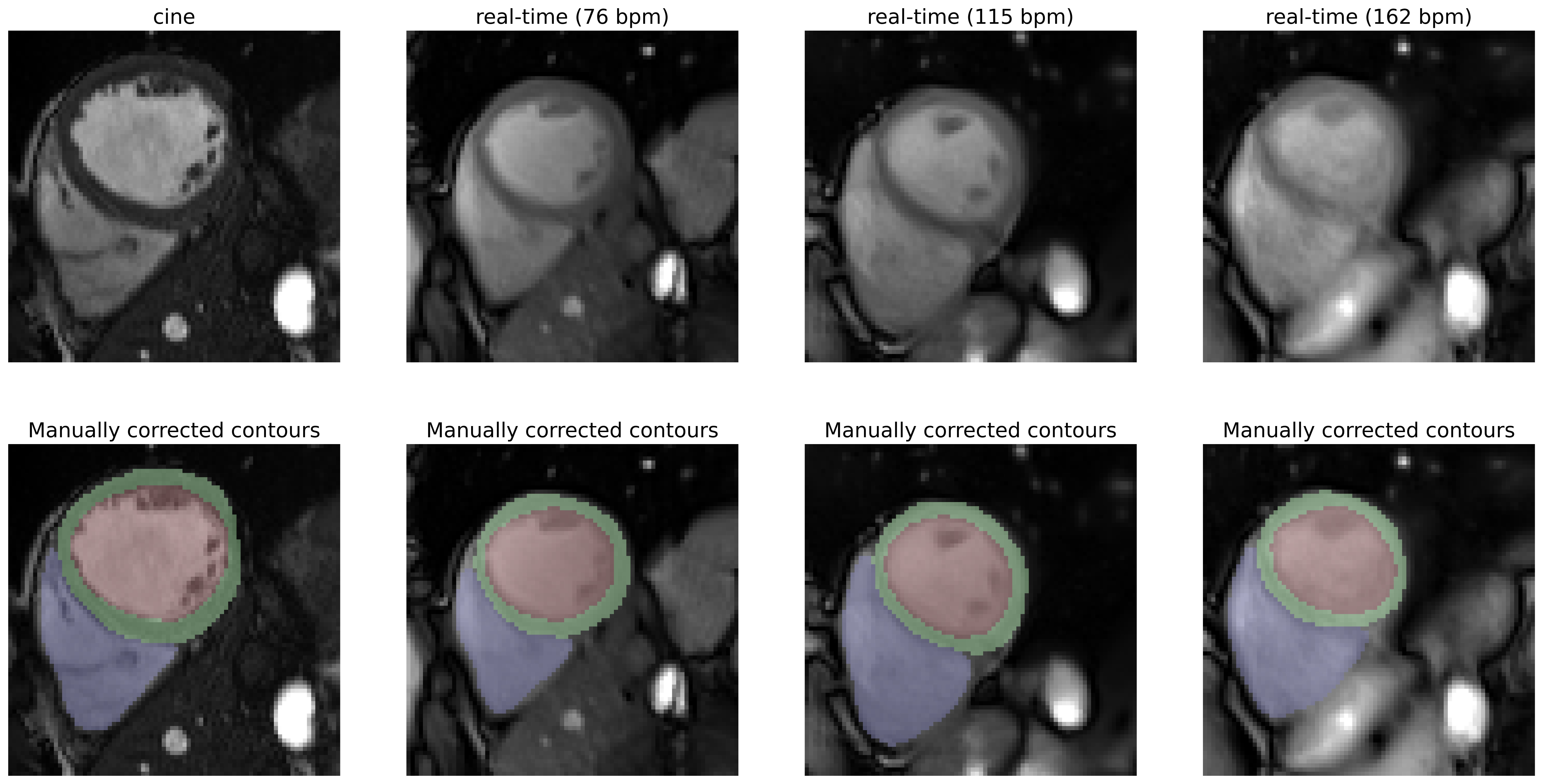}
	\caption[Comparison of different measurement types]
	{(\textbf{Top}) Mid-ventricular slices in ED phase of the same volunteer for cine and real-time free-breathing at different heart rates
	and (\textbf{bottom}) corresponding manually corrected segmentation in a short axis view are shown.
	Image quality decreases and reconstruction artifacts increase with an increasing heart rate.
	The left ventricular endocard (red), the left ventricular myocardium (green), and the right ventricle (blue) are segmented.}
	\label{fig:measurementtypes}
\end{figure*}

A comparison of cine and real-time CMR measurements and representative manually corrected contours are presented in Figure \ref{fig:measurementtypes}.

\subsection{Dataset}

The dataset consists of cine and real-time images
from 15 healthy volunteers (7 male; 8 female; aged 55$\pm$8 (s.d.))
from a consecutive series of exams performed
at the Institute for Diagnostic and Interventional Radiology of the University Medical Center G\"ottingen.
It is part of a larger dataset acquired in a previous research study on real-time MRI with the approval of the local ethics committee.
Consent for publication was obtained from all participants in the study.

All volunteers were measured with the same protocol.
The healthy volunteers underwent CMR in supine position using a
32-channel cardiac surface receiver coil at 3\,T (Skyra, Siemens Healthineers, Germany).
Conventional imaging at rest included a balanced steady-state free precession (bSSFP) ECG-gated cine sequence
to create a short-axis stack covering the entire heart, including both ventricles and atria.
Real-time CMR data acquisition was performed during free-breathing and without ECG-synchronization
at rest and under two different levels of exercise stress.
A detailed summary of acquisition parameters of cine and real-time CMR can be found in Supplementary Table \ref{tab:acq_param}.

Real-time acquisiton was based on a bSSFP sequence using a highly undersampled
radial encoding scheme and iterative image reconstruction \cite{Uecker_NMRBiomed._2010,Zhang_J.Cardiov.Magn.Reson._2010}.
Exercise stress was introduced using a CMR-compatible ergometer in the supine position (Lode, Leiden, Netherlands),
as previously described in \cite{Steinmetz_Circ.Cardiovasc.Imaging_2021}.

After performing real-time free-breathing measurements at rest (RT rest),
exercise stress was increased until a target heart rate of 110 beats per minute (bpm) was reached.
Measurements at this target heart rate are referred to as real-time stress (RT stress).
Finally, exercise stress was increased to the subjective, maximal exercise stress of each person and a measurement was performed (RT max stress).
All persons gave written informed consent before each CMR examination.

\subsection{Segmentation}

\subsubsection{Reference}

To create a reference segmentation, manually corrected segmentation was performed
using Medis (version 4.0.56.4, QMass® 8.1, Medical Imaging Systems, Leiden, Netherlands).
Contours for the left ventricular endocardium,
the left ventricular epicardium, and the right ventricle were created on short axis slices in end-diastole and end-systole.
The segmentation of the left ventricular myocardium is formed by the area between the left ventricular epicardium and endocardium.
Contours automatically created by the software were then manually corrected to create the reference
in accordance with the standard procedure used in clinical routine.
Contour creation followed the contour protocol used for the Automated Cardiac Diagnosis Challenge (ACDC) \cite{Bernard_IEEETrans.Med.Imag._2018}:
Papillary muscles are included in the cavity and LV endocardial contours follow the limit defined by the aortic valve at the base of the LV.

For real-time measurements at rest and under physical stress, images in ED and ES phase in the whole time series (120-150 images) were segmented.
At rest, the time series spans 3-4 heartbeats. Under physical stress, the time series spans 6-9 heartbeats.
For real-time measurements at maximal physical stress, only images in ED and ES phase among the first 50 images of the time series (2-4 heartbeats) were segmented.
For the measurements of three volunteers at maximal physical stress, image quality was too poor to create reasonable reference contours.

\subsubsection{Neural networks for automatic segmentation}

This work evaluates two different neural networks for automatic segmentation.
The first is the automatic contour creation from the commercial software Medis DL ACD (Medis deep learning
automatic contour detection) in QMass 8.1. We refer to this method as comDL.

It should be stressed that comDL is intended to be used for cine CMR as a basis
for the contours, to be manually corrected and checked as described above.
Since manual correction is generally not feasible for the high number of images
acquired using real-time MRI, the scope of this work is to evaluate the performance of deep-learning methods for automatic
segmentation.

The second neural network evaluated in this work is nnU-Net.
It is freely available and offers trained weights for various image segmentation challenges in the medical field.
The model has already shown great versatility and was successfully used for a variety of medical segmentation tasks, e.g.
achieving first place for all segmentation classes in the cardiac segmentation challenge "Automated Cardiac Segmentation Challenge" (ACDC) \cite{Isensee_Nat.Methods_2021}.

Preliminary analysis (results not shown) showed that nnU-Net performs better if applied on
individual 2D images rather than on a stack of 2D images, e.g. a time series of
real-time MRI in a single slice or a stack of cine images within the same cardiac phase.
The benefit of independent normalization of each image
increased for more challenging segmentation tasks (see Supplementary Table \ref{tab:nnunet_single_stack}). 
The application of 2D nnU-Net was identified as the best configuration for this task
when compared with 3D nnU-Net and the ensemble of both models (see Supplementary Table \ref{tab:nnunet_ensemble}).

Consequently, the 2D version of nnU-Net with weights pre-trained on the ACDC dataset was applied
on single images of the dataset for all cine and real-time measurements.

In March 2023, a new version of nnU-Net was published (nnU-Net V2). As of 2024-01-19, no pre-trained weights have been published for the second version
and segmentation performance reportedly remains the same \cite{Isensee_git_2023}.
As such, this work evaluates the first version of nnU-Net.

\subsubsection{Evaluation criterion}

It evaluates the overlap of a predicted segmentation of a neural network $P_k$ with a reference segmentation $R_k$ for each individual segmentation
class $k$ and is defined as
\begin{equation}
	\mathrm{DC} = 2 \frac{P_k \cap R_k}{P_k + R_k}.
\end{equation}
DC is a value between 0 and 1, with 0 denoting no overlap between prediction and reference and 1 denoting perfect agreement.

\subsection{Calculation of heart rates}\label{section:calc_heart_rates}

We also categorized all
real-time images based on heart rates, as these differed for RT max stress.
For the calculation of heart rates, we used the three central slices of all slices between the base and apex.
In the central slices, the left ventricle is present during the entire time series despite respiratory motion,
which can cause a displacement of the heart in and out of the imaging plane in basal and apical slices.
Heartbeats per minute (bpm) are calculated based on the duration between two end-diastolic phases.

\subsection{Cardiac function parameters}

End-diastolic volume (EDV), end-systolic volume (ESV), and ejection fraction (EF) were computed.
To minimize the influence of respiratory motion on EDV and ESV, images in the ED and ES phase of the cardiac cycle during end-expiration were
manually selected for each slice.
For comDL and nnU-Net, cardiac function parameters were calculated based on segmentation of the same selected images.
Ventricular volumes were calculated with Simpson's rule \cite{Greenberg_J.Thorac.Imag._2000,Mahnken_Rofo-Fortschr.Rontg._2004}.

For intra-observer variability, manually corrected contours for the derivation of the clinical measures of all volunteers were created
three to six months after the initial segmentation.
For inter-observer variability, manually corrected contours for the derivation of the clinical parameters were created
for the first five volunteers by a second reader with experience in cardiac segmentation.
Single images in the ED and ES phase during end-expiration were once again chosen from each slice and
EDV, ESV, and EF were derived from newly created, manually corrected contours.

Previously reported \cite{Bai_J.Cardiov.Magn.Reson._2018} inter-observer variability was chosen as an additional reference for comparison
because it represents the accuracy of the evaluation in the clinical workflow of cine CMR.
Measurements are usually evaluated by a single person and the accuracy of the method is determined by the variance between different human observers.

\subsection{Statistics}

We evaluated cardiac function parameters through Bland-Altman plots and paired two-sample t-tests.
The cardiac function parameters EDV, ESV, and EF that were derived from comDL and nnU-Net segmentations were compared to
the respective values derived from manually corrected contours (see Supplementary Fig. \ref{fig:BA_CF_nnunet} and \ref{fig:BA_CF_comDL}).
The comparison includes cine CMR and real-time CMR measurements at rest and under stress of all volunteers.
Additionally, cardiac function parameters of cine CMR and real-time CMR at rest were compared with each other (see Supplementary Fig. \ref{fig:BA_CF_cine_rt}).
T-tests were performed under the null hypothesis with a significance level of $\alpha = 0.05$.

\section{Results}

\subsection{Segmentation accuracy}

The data for all 15 volunteers could be successfully analyzed for cine CMR and real-time free-breathing CMR
at rest and under exercise stress.
For the measurements of three volunteers at maximal physical stress,
image quality was too poor to create reasonable reference contours and
thus only data of 12 volunteers were analyzed.

\begin{figure*}[ht!]
	\centering
	\includegraphics[width=\textwidth]{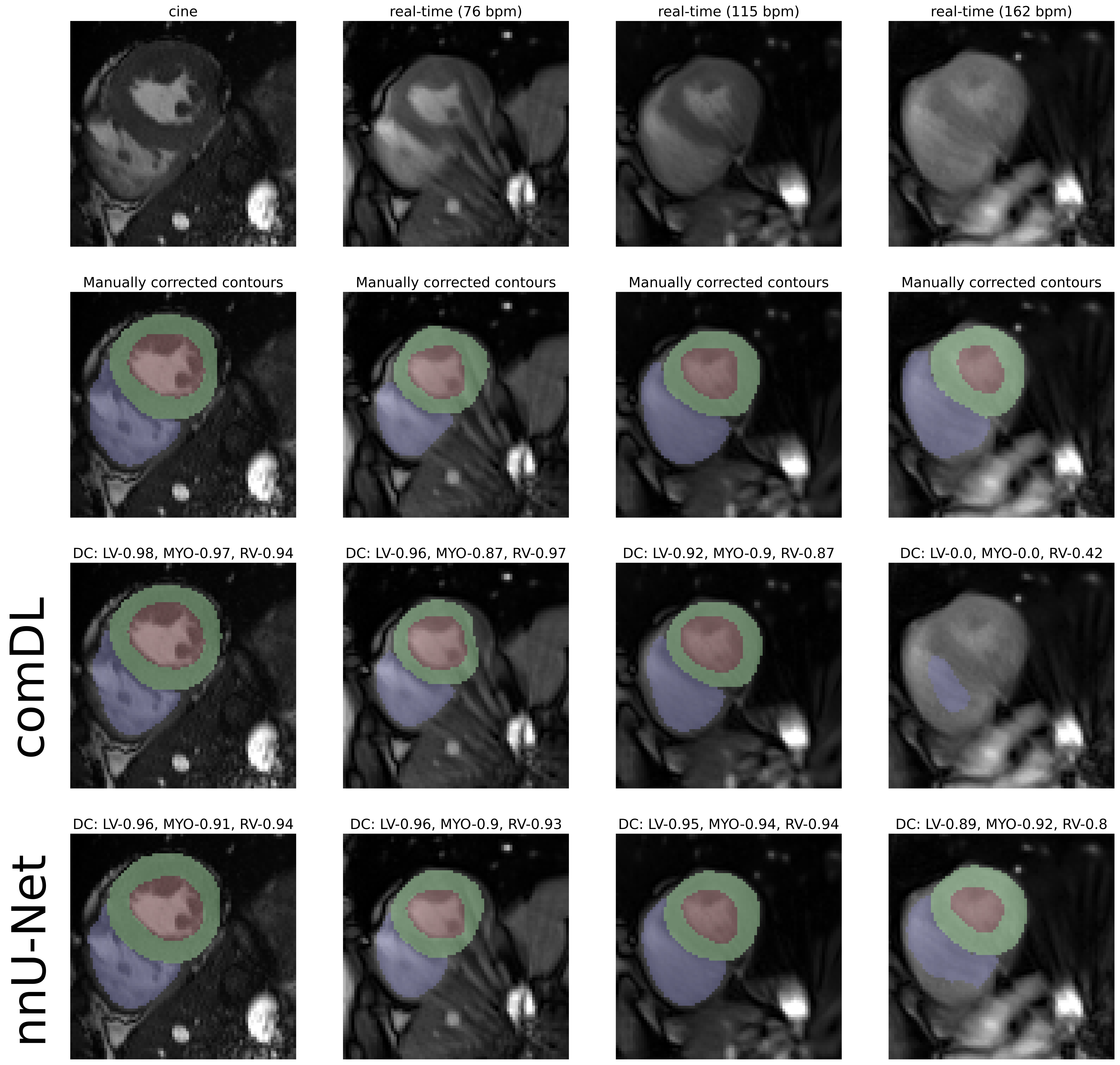}
	\caption[Representative segmentations of manually corrected contours and deep learning methods]
	{Representative segmentations of manually corrected contours and deep learning methods.
	(\textbf{First row}) Mid-ventricular slices in ES phase of a volunteer for cine and real-time free-breathing at different heart rates
	with (\textbf{second row}) corresponding manually corrected, (\textbf{third row}) comDL, and (\textbf{fourth row}) nnU-Net segmentation.
	Accuracy of segmentation is measured with Dice's coefficient (DC). DC for left ventricular endocard (LV), left ventricular myocardium (MYO),
	and right ventricle (RV) are given for each segmentation.}
	\label{fig:dicecoefficient}
\end{figure*}

For cine CMR, the DC values for both neural network segmentations are comparable (Table \ref{tab:nnunet_comDL_cine_rt})
and show a high correlation with conventional segmentation.
Figure \ref{fig:dicecoefficient} shows representative DC cases.

\begin{table}[ht]
	\caption[Dice's coefficients of deep learning segmentation for cine and real-time CMR]
	{Dice's coefficients of nnU-Net and comDL segmentation for cine and real-time CMR.
	The table features the mean and standard deviation (in parenthesis) of the
	Dice's coefficients for LV, MYO, and RV for all volunteers.
	For cine CMR, images were separated into ED and ES phase.
	For RT max stress, only data of 12 volunteers were analyzed, because in three cases the image quality was too poor
	to create reasonable reference contours.
	Mean and standard deviation (in parenthesis) of previously reported values for inter-observer variability from three different human observers
	for cine CMR are shown for comparison.}
	\label{tab:nnunet_comDL_cine_rt}
	\begin{tabular*}{\textwidth}{@{\extracolsep\fill}lccc}
		\hline
		n=15 & LV & MYO & RV \\
		\hline
		nnU-Net cine ED		& 0.98	(0.00)	& 0.91	(0.02)	& 0.92	(0.05)	\\
		nnU-Net cine ES		& 0.92	(0.04)	& 0.91	(0.03)	& 0.88	(0.06)	\\
		\hline
		nnU-Net cine		& 0.95	(0.02)	& 0.91	(0.02)	& 0.90	(0.03)	\\
		nnU-Net RT		& 0.94	(0.02)	& 0.89	(0.02)	& 0.90	(0.03)	\\
		nnU-Net RT stress	& 0.92	(0.03)	& 0.85	(0.03)	& 0.83	(0.11)	\\
		nnU-Net RT max stress (n=12)	& 0.91	(0.03)	& 0.83	(0.04)	& 0.79	(0.16)	\\
		\hline
		\hline
		comDL cine ED		& 0.98	(0.02)	& 0.97	(0.02)	& 0.92	(0.06)	\\
		comDL cine ES		& 0.95	(0.05)	& 0.95	(0.04)	& 0.88	(0.08)	\\
		\hline
		comDL cine		& 0.97	(0.03)	& 0.96	(0.02)	& 0.90	(0.06)	\\
		comDL RT		& 0.93	(0.04)	& 0.88	(0.05)	& 0.92	(0.05)	\\
		comDL RT stress		& 0.79	(0.15)	& 0.72	(0.15)	& 0.79	(0.14)	\\
		comDL RT max stress  (n=12)	& 0.70	(0.21)	& 0.62	(0.19)	& 0.69	(0.18)	\\
		\hline
		\hline
			& 0.94 (0.04)	& 0.88 (0.02)	& 0.87 (0.06)	\\
		inter-observer cine (n=50) \cite{Bai_J.Cardiov.Magn.Reson._2018}
			& 0.92 (0.04)	& 0.87 (0.03)	& 0.88 (0.05)	\\
			& 0.93 (0.04)	& 0.88 (0.02)	& 0.89 (0.05)	\\
		\\
		\hline
	\end{tabular*}
\end{table}
\newpage
For cine CMR, the nnU-Net performance (Table \ref{tab:nnunet_comDL_cine_rt}) is comparable to the results achieved
on the ACDC test dataset \cite{Isensee_Nat.Methods_2021}.
The Dice's coefficients reported here are slightly higher, as might be expected for the application on healthy subjects
as compared to different pathology groups in the ACDC test dataset.

Real-time free-breathing measurements
can be categorized based on the form of acquisition, e.g. at rest (RT rest),
under a level of stress selected according to a targeted heart rate of 110\,bpm (RT stress) and maximal exercise stress (RT max stress),
for which the stress level and heart rate varies by volunteer.
\begin{figure*}[h!]
	\centering
	\includegraphics[width=\textwidth]{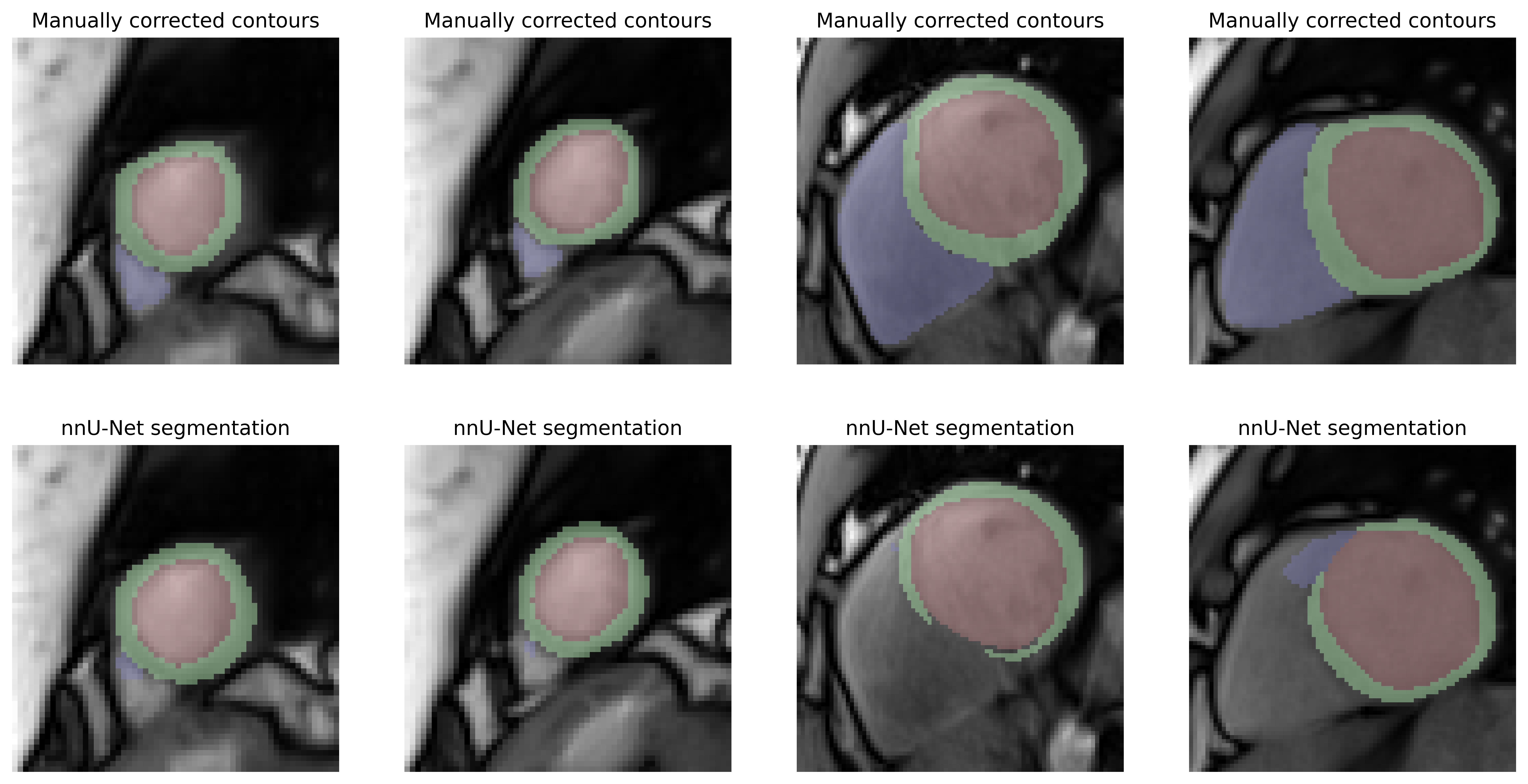}
	\caption[Example segmentation failures of nnU-Net for real-time CMR under exercise stress]
	{Example segmentation failures of nnU-Net for real-time CMR under exercise stress.
	Incomplete segmentation of the right ventricle in the apical region (first and second column).
	Anatomically incoherent segmentation of the myocardium and right ventricle in the basal region (third and fourth column).}
	\label{fig:limits}
\end{figure*}
Mean heart rate and standard deviation for all volunteers can be found as Supplementary Table \ref{tab:vol_heart_rate}.
The different real-time free-breathing CMR measurements fall into different heart rates spans,
RT - 55 to 77\,bpm, RT stress - 107 to 120\,bpm, RT max stress - 121 to 164\,bpm.

The accuracies of nnU-Net and comDL segmentation based on the heart rate categorization are presented in Table \ref{tab:nnunet_comDL_cine_rt}.
The nnU-Net model shows good generalizability with high accuracies of segmentation for both real-time CMR (DC: LV 0.94, MYO 0.89, RV 0.90)
and real-time stress (DC: LV 0.92, MYO 0.85, RV 0.83).
The accuracies for real-time CMR at rest are in the order of previously reported \cite{Bai_J.Cardiov.Magn.Reson._2018} inter-observer variability for cine CMR.
With the exception of the RV segmentation for real-time CMR at rest, the accuracy of nnU-Net predictions exceeds the accuracy of comDL
in real-time CMR.

Example failure cases for the nnU-Net are shown in Fig. \ref{fig:limits}.
Images in the basal and apical region of the heart were especially prone to segmentation failures.

\begin{figure}[ht]
	\centering
	\includegraphics[width=\textwidth]{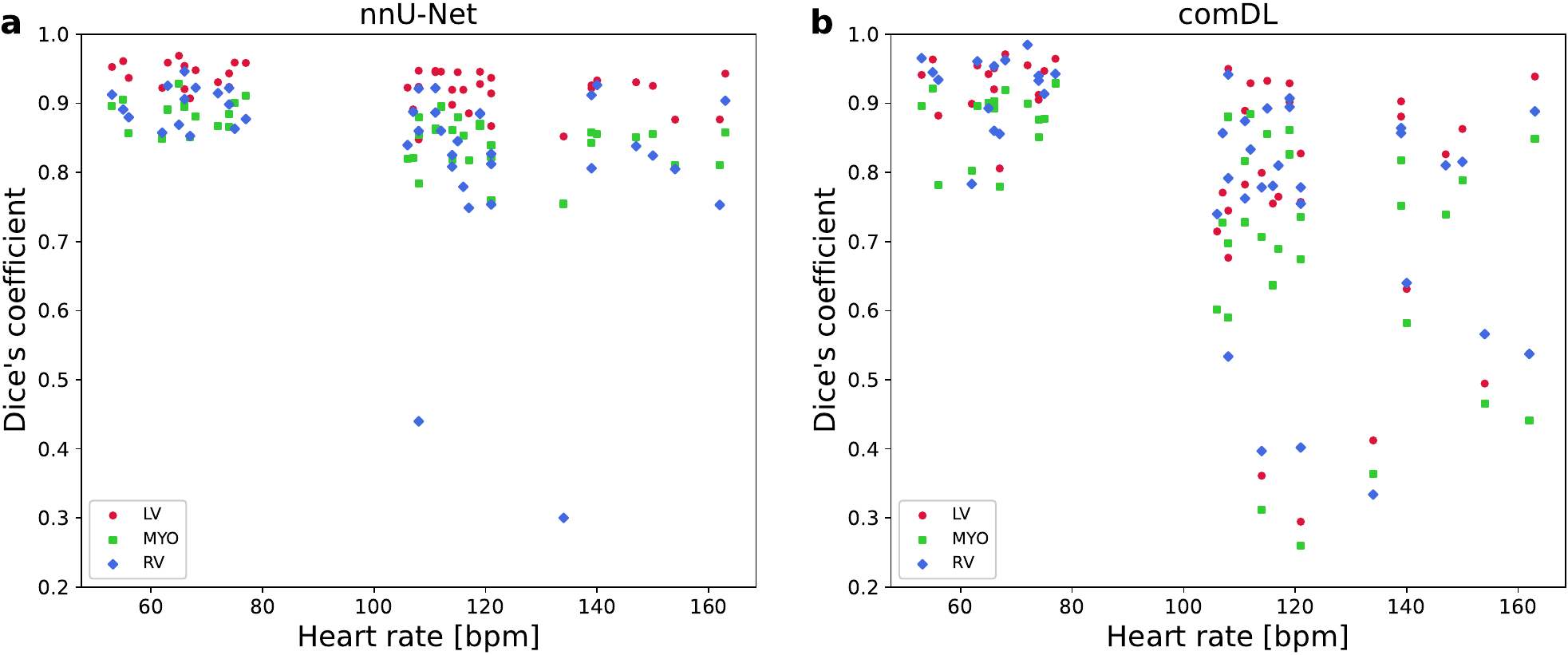}
	\caption[Dice's coefficient of real-time CMR measurements plotted against heart rate]
	{Dice's coefficient of nnU-Net and comDL segmentation for real-time CMR measurements plotted against heart rate.
	DC values of LV, MYO, and RV are calculated for (\textbf{a}) nnU-Net and (\textbf{b}) comDL segmentation in respect to manually corrected contours.
	Each data point presents the average DC of a segmentation class for a single real-time measurement of a volunteer.
	Real-time CMR at rest, under exercise stress and maximal exercise stress are presented by their average calculated heart rate.}
	\label{fig:DC_vs_bpm}
\end{figure}

To better visualize the relationship between heart rate and the neural network segmentation accuracy,
Fig. \ref{fig:DC_vs_bpm} shows the nnU-Net and comDL Dice's coefficients by calculated heart rate for real-time CMR.
Two outliers are visible for the RV segmentation of the nnU-Net.
Both data points correspond to the RT and RT stress measurements of the same volunteer.
While the RV segmentation performs well for cine and RT it nearly completely fails for RT stress and RT maxstress,
most likely because of smooth transitions between RV and neighboring tissue.
A representative comparison can be found as Supplementary Fig. \ref{fig:RV_outlier}.

\subsection{Cardiac function parameters}

For cine and real-time CMR, the EDV, ESV, and EF were derived from the
manually corrected contours for manually selected images in the ED and ES phases.
The absolute values for each volunteer can be found in Supplementary Table \ref{tab:vol_cardiac_function}. 
The comparison between these values and those derived from the nnU-Net and comDL segmentation
is presented in Table \ref{tab:nnunet_comDL_EDV_ESV_EF} in form of absolute and relative differences.

\begin{table}[h!]
	\caption[Cardiac function parameters derived from comDL and nnU-Net]
	{Cardiac function parameters derived from comDL and nnU-Net segmentation compared to reference values obtained with manually corrected segmentation.
	The table features the differences of EDV, ESV, and EF of cine and real-time CMR of all volunteers.
	The mean and standard deviation (in parenthesis) of the difference are reported for (\textbf{a}) the absolute and (\textbf{b}) the relative difference.
	Intra- and inter-observer variability are given for values which were derived from manually corrected contours of newly selected images.
	Previously reported values for inter-observer variability from three different human observers for cine CMR are presented for comparison.}
	\label{tab:nnunet_comDL_EDV_ESV_EF}
	\begin{tabular*}{\textwidth}{@{\extracolsep\fill}lccc}
	\hline
	(\textbf{a}) Absolute difference & & & \\
	n=15 & EDV [mL] & ESV [mL] & EF [\%]\\
	\hline
	nnU-Net cine		& 1.8 (1.5)	& 3.7 (2.9)	& 2.5 (2.2)	\\
	nnU-Net RT		& 2.9 (2.1)	& 3.5 (3.6)	& 2.6 (2.6)	\\
	nnU-Net RT stress	& 11.4 (16.3)	& 2.9 (1.8)	& 3.6 (4.5)	\\
	\hline
	comDL cine		& 1.0 (0.7)	& 3.0 (2.7)	& 1.8 (1.4)	\\
	comDL RT		& 6.0 (5.4)	& 5.4 (4.9)	& 4.0 (5.4)	\\
	comDL RT stress		& 32.7 (28.5)	& 7.3 (5.0)	& 26.3 (47.0)	\\
	\hline
	intra-observer cine		& 2.8 (3.9)	& 1.4 (1.9)	& 1.5 (1.6)	\\
	intra-observer RT		& 3.5 (4.2)	& 4.5 (4.5)	& 2.9 (2.3) 	\\
	intra-observer RT stress	& 4.1 (3.3)	& 4.3 (3.9)	& 3.5 (2.2) 	\\
	\hline	\hline
	inter-observer cine (n=5)	& 4.4 (4.9)	& 8.5 (7.9)	& 4.5 (3.4)	\\
	inter-observer RT (n=5)		& 4.8 (3.1)	& 11.1 (7.6)	& 7.1 (4.8) 	\\
	inter-observer RT stress (n=5)	& 11.5 (6.6)	& 9.4 (8.9)	& 5.1 (5.1) 	\\
	\hline
	\hline
		& 6.1 (4.4)	& 4.1 (4.2)	& 3.1 (2.1)	\\
	inter-observer cine (n=50) \cite{Bai_J.Cardiov.Magn.Reson._2018}
		& 8.8 (4.8)	& 6.7 (4.2)	& 3.0 (2.4)	\\
		& 4.8 (3.1)	& 7.1 (3.8)	& 3.8 (1.8)	\\
	\\
	(\textbf{b}) Relative difference & & & \\
	n=15 & EDV [\%] & ESV [\%] & EF [\%]\\

	\hline
	nnU-Net cine		& 1.2 (1.0)	& 6.7 (5.2) 	& 4.2 (3.9) 	\\
	nnU-Net RT		& 2.4 (2.4)	& 6.6 (10.1)	& 4.6 (4.2)	\\
	nnU-Net RT stress	& 7.2 (9.0) 	& 5.5 (3.6)	& 5.9 (6.8)	\\
	\hline
	comDL cine		& 0.7 (0.5)	& 5.2 (3.9) 	& 2.9 (2.4) 	\\
	comDL RT		& 4.7 (6.1)	& 8.1 (5.8)	& 7.3 (10.2)	\\
	comDL RT stress		& 22.8 (22.9)	& 14.2 (11.6)	& 42.1 (75.9)	\\
	\hline
	intra-observer cine		& 1.7 (2.3)	& 2.7 (3.8)	& 2.5 (2.7)	\\
	intra-observer RT		& 2.4 (2.6)	& 6.3 (5.2)	& 5.2 (4.2)	\\
	intra-observer RT stress	& 2.9 (2.5)	& 7.6 (5.4)	& 5.9 (4.1)	\\
	\hline
	inter-observer cine (n=5)	& 3.2 (3.6)	& 15.9 (12.4)	& 7.1 (5.4)	\\
	inter-observer RT (n=5)		& 3.1 (1.8)	& 17.5 (10.5)	& 11.9 (8.5) 	\\
	inter-observer RT stress (n=5)	& 8.3 (4.5)	& 16.9 (14.5)	& 9.2 (10.0) 	\\
	\hline
	\hline
		& 4.2 (3.1)	& 6.8 (7.5)	& 5.1 (3.7)	\\
	inter-observer cine (n=50) \cite{Bai_J.Cardiov.Magn.Reson._2018}
		& 6.3 (3.3)	& 12.5 (8.5) 	& 4.9 (3.8)	\\
		& 3.4 (2.2)	& 11.7 (5.1)	& 6.6 (3.2)	\\
	\hline
	\end{tabular*}
\end{table}

For cine CMR, manual corrections of comDL led to only slight differences in EDV and ESV.
For real-time CMR, the difference between the nnU-Net segmentation and manually corrected contours compares well to intra- and inter-observer variability for real-time CMR
as well as to
the inter-observer \cite{Bai_J.Cardiov.Magn.Reson._2018} variability previously reported for cine CMR (Table \ref{tab:nnunet_comDL_EDV_ESV_EF}).
For nnU-Net, mean differences of 2.4\,\% for EDV, 6.6\,\% for ESV, and 4.6\,\% for EF were obtained for real-time CMR at rest.
For nnU-Net, the absolute and relative differences of EDV, ESV, and EF are less than the inter-observer variability,
although the direct comparability is limited because images were newly selected for intra- and inter-observer variability, while
the nnU-Net segmentation has been compared to the manually corrected contours of the same images.
The different selection process of the images may influence the variability a lot more
than a different delineation of LV within the same image.
Relative intra-observer variability increases from cine CMR to real-time CMR at rest, and again to real-time CMR at stress.
Relative intra- and inter-observer variability is overall higher for ESV than for EDV, which is in agreement with previously reported inter-observer variability.
Inter-observer variability is slightly higher for ESV in cine CMR than what was previously reported but only increases somewhat for RT and RT stress.

Clinical measures derived with nnU-Net and comDL are compared to references in form of Bland-Altman plots (Supplementary Fig. \ref{fig:BA_CF_nnunet} and \ref{fig:BA_CF_comDL}).
According to paired two-sample t-tests, differences between nnU-Net and manually corrected contours are significant
for ESV and EF for cine CMR ($P<0.001$)
and not significant for EDV ($P=0.23$) for cine CMR and for all cardiac function parameters for RT and RT stress ($P>0.5$).
Differences between comDL and manually corrected contours are not significant for EDV for cine ($P=0.16$) and real-time CMR ($P=0.61$)
and significant for all other cardiac function parameters of cine, RT and RT stress ($P<0.05$).
Individual $P$ values are shown in the Bland-Altman plots.

Although a comparison of cardiac function parameters between cine and real-time CMR was not the focus of this work,
we note that the data show a higher mean EDV (mean difference 8.9\,mL) and a lower mean ESV (mean difference -4.1\,mL)
for cine CMR (see Supplementary Fig. \ref{fig:BA_CF_cine_rt}).
These deviations are in the order of previously reported \cite{Feng_Magn.Reson.Med._2013} differences between cine and real-time CMR.
According to paired two-sample t-tests, differences between cine and real-time CMR are significant ($P<0.05$)
for all cardiac function parameters for manually corrected contours. 

\section{Discussion}

In this study of cine CMR and real-time free-breathing CMR at rest and under exercise stress of 15 healthy volunteers,
we found that the accuracy of comDL and nnU-Net is in the order of inter-observer variability for cine CMR and real-time CMR at rest.
Consequently, both methods are viable for a prospective, automated evaluation with little or no manual correction.
For real-time CMR under exercise stress, nnU-Net is usable as a basis for manually corrected contours.

In our study, comDL contours are the basis for manually corrected contours, which act as reference.
These automatically created contours require little manual correction, confirming the good performance of comDL for cine CMR.
However, our study showed comDL to be less accurate than nnU-Net for real-time CMR, especially for RT stress
and RT max stress. These results might be expected from the perspective that comDL was designed and most likely optimized for cine CMR,
for which it performed very well.

nnU-Net shows better generalizability for real-time CMR than comDL,
having its segmentation accuracy decrease less between real-time and exercise stress measurements.
For measurements under exercise stress, nnU-Net may have reached the limit of its applicability for fully automatic segmentation.
Although its performance remains quite good, as demonstrated by high Dice's coefficients and a low mean absolute difference in ESV,
the mean difference in EDV is significantly larger than intra-observer variability.
While the accuracy of nnU-Net might not yet be sufficient for fully automatic segmentation,
it shows promising results for an increased degree of automation in the future.

In this work, we observed the highest degree and frequency of deviations between reference and neural network segmentation in the basal and apical regions of the heart.
These regions were also identified as the most problematic factor for neural network segmentation
in the evaluation of ACDC \cite{Bernard_IEEETrans.Med.Imag._2018}.
Due to the ambiguity of slice positions, which can still include or exclude certain segmentation classes,
comDL and nnU-Net showed some difficulty in correctly segmenting these areas.
One approach to this issue is the usage of multiple neural networks individually trained on specific heart regions \cite{Amirrajab_Lect.Notes.Comput.Sc._2022}.
This however requires manually labeling input data prior to segmentation.

One previous study concerning the segmentation of real-time free-breathing CMR with deep learning neural networks is the work by Yang et al.
\cite{Yang_BioMedRes.Int._2019}, which created a custom neural network based on the U-Net \cite{Ronneberger_Lect.NotesComput.Sc._2015} architecture
and also trained on the ACDC dataset. They evaluated their network on end-systolic and end-diastolic phases in the end-expiration state of healthy volunteers.
Our results for real-time free-breathing CMR show a higher segmentation accuracy for
comDL and nnU-Net compared to results of \cite{Yang_BioMedRes.Int._2019} (DC: LV 0.919, MYO 0.806, RV 0.818),
showing the progress of segmentation networks for real-time CMR.

New methods have been developed for cardiac segmentation in recent years,
e.g. the usage of transformers within neural networks for segmentation is a more recent idea than the use of
convolutional neural networks and might be an essential element for future research \cite{El-Taraboulsi_ArtificialIntelligenceintheLifeSciences_2023}.
A combination of nnU-Net with transformers \cite{Zhou_IEEETrans.ImageProcessing_2023} shows promising results, especially in an ensemble with the unmodified nnU-Net.
However, the highest performance of deep learning neural networks for segmentation in the field of CMR is still achieved
by convolutional neural networks, often based on the U-Net or the nnU-Net \cite{El-Taraboulsi_ArtificialIntelligenceintheLifeSciences_2023}.
Therefore, it seems reasonable to have nnU-Net as the state-of-the-art method, in particular because of its accessibility through pre-trained weights.

The development and application of deep learning methods is an active research topic in radiology \cite{Hosny_Nat.Rev.Cancer_2018,ESR_InsightsImaging_2019,Reardon_Nature_2019}.
Standards for the reporting of artificial intelligence methods were established for the medical field as a whole in \cite{Mongan_Radiol.Artif.Intell._2020}
and the topic was specifically discussed for CMR in \cite{Alabed_Front.Cardiovasc.Med._2022,Maiter_Front.Radiol._2023}.

For cardiac segmentation, deep learning methods have already been successfully
integrated into clinical routines and have substantially reduced the number of
manual corrections needed. With recent progress, a fully automatic workflow seems feasible even
for real-time CMR, but essential steps are still missing.
If the evaluation of real-time free-breathing measurements should work analogously to cine CMR,
images within the same respiratory motion state must be evaluated across slices.
The identification of the respiratory state would need to be automated, e.g. with
the help of an external device like a respiratory belt, or by automatic analysis
of the images.
Based on the respiratory motion state, images in the correct cardiac phase can then
be automatically selected based on LV area.
To keep manual corrections to a minimum, additional priors in the form of confidence maps, such as uncertainty maps \cite{Sander_Proc.SPIE_2019},
could be used to quantify the need of manual correction.
For patients with arrhythmia, arrhythmic heartbeats would need to be distinguished automatically
from regular (sinus) rhythm.

Some limitations of our study must be noted.
Firstly, extrasystolic heartbeats within the selected images were not excluded, as the limited duration of the time series did not always allow
the monitoring of prior and subsequent heartbeats to fully exclude an extrasystole.
However, this only affects the validity of the resulting cardiac function parameters,
not the comparison between the manually corrected contours and the deep learning methods, as the same images have been evaluated for all methods.
Secondly, only a relatively small number of healthy volunteers were included in this study
and the validity for clinical application on patients still needs to be demonstrated.
Thirdly, no detailed methodological information can be provided for comDL, as it is part of a commercial software.
It still serves as a reference to the clinical standard.
Finally, only two deep-learning methods have been compared,
which limits the generalizability of the results to machine learning methods in general.
We hope to address this with the publication of image data and
code to enable the testing and comparison of other deep-learning methods on the same data.

\section{Conclusion}

In this study, we assessed the feasibility of automatic cardiac segmentation on real-time CMR using
deep learning methods.
Two deep learning methods originally designed or trained for segmentation of cine CMR
were evaluated for cine and real-time MRI in comparison to a manually corrected reference segmentation.
The segmentation accuracy is superior in cine CMR compared to real-time CMR at rest and diminishes further for
real-time CMR under exercise stress.
The accuracy for real-time CMR at rest is in the range of reported inter-observer variability of cine CMR for both networks.
In this work, comDL shows very good performance for segmentation on cine CMR but less applicability for real-time CMR compared to nnU-Net.
For real-time CMR at rest, cardiac function parameters obtained with nnU-Net
segmentation are in the range of intra-observer variability.
For real-time CMR under exercise stress, the performance of the deep learning
methods - while still not sufficient for a fully automatic segmentation -
is promising.

\section*{Acknowledgements}

This work was funded by the "Nieders\"achsisches Vorab" funding line of the Volkswagen Foundation.
This work was supported by the DZHK (German Centre for Cardiovascular Research).
We are particularly grateful for the assistance given by Florian Kohler
who recruited volunteers and performed the scans together with Ulrike
K\"ochermann and Tanja Otto.

\section*{Author contributions statement}

MS experimental analysis, interpretation of data, and writing of manuscript;
CUB design of the MRI study, experimental workup, volunteer recruitment, and revision of the manuscript;
JL revision of the manuscript;
MU development of real-time MRI methods, data analysis,
and revision of the manuscript.

\subsection*{Competing interests}
MU is a co-inventor of a patent covering the real-time MRI
technique used in this study.
All other authors hereby state that they have no financial or personal relationships
with other people or organizations that could inappropriately influence or bias their work.

\subsection*{Availability of data and materials}
Cine and real-time CMR images, comDL and manually corrected contours, nnU-Net segmentation, and manually selected end-expiration indexes
are available at Zenodo under \url{https://zenodo.org/records/10117943} (DOI:10.5281/zenodo.10117943).

\subsection*{Code availability}
Code for the reproduction of figures and results is available under \url{https://github.com/mrirecon/dl-segmentation-realtime-cmr}.

\clearpage
\appendix
\section*{Supporting Information}

\begin{figure*}[ht]
	\centering
	\includegraphics[width=\textwidth]{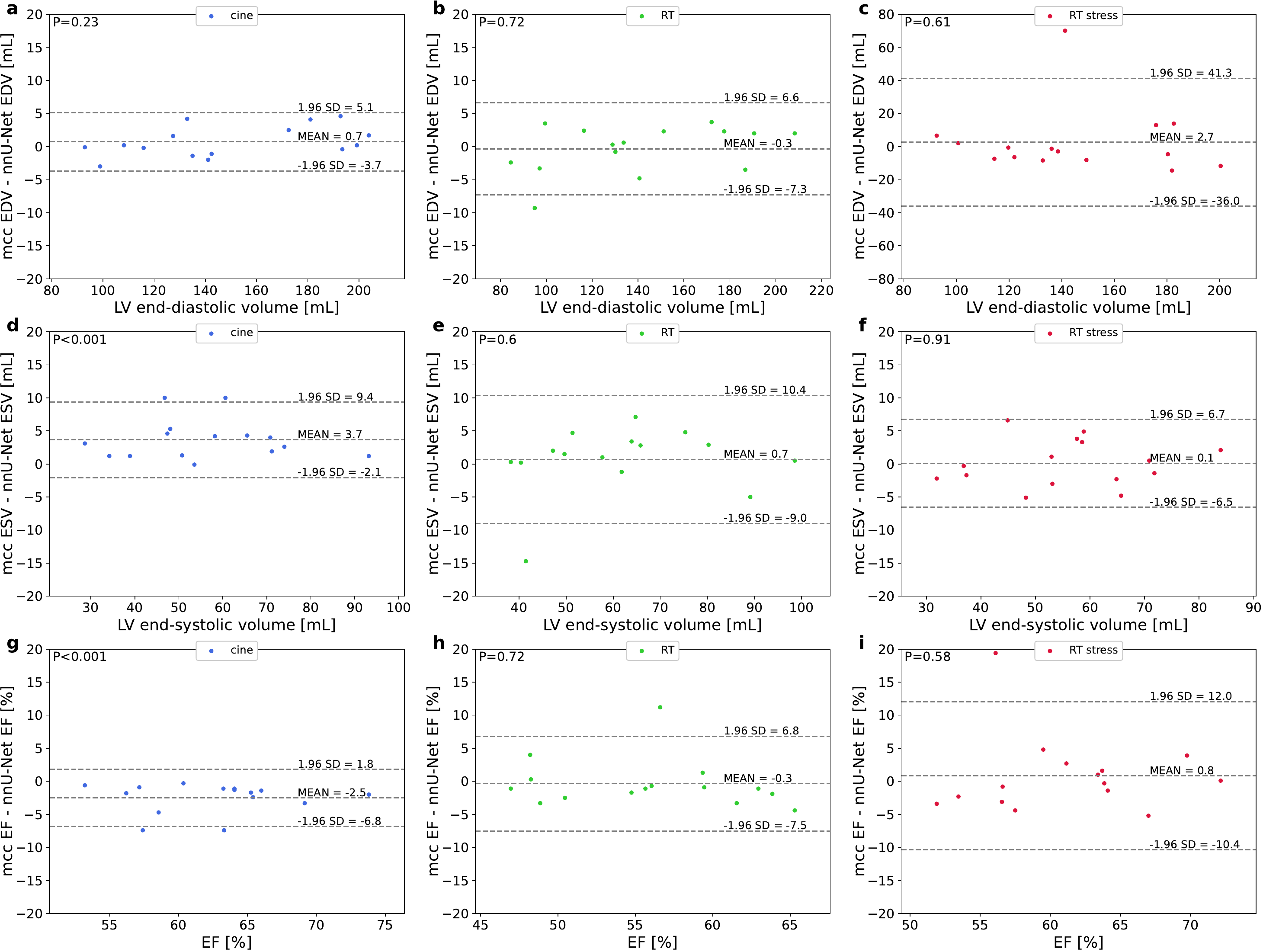}
	\caption{Bland-Altman plots for cardiac function parameters of nnU-Net.
	The cardiac function parameters of (\textbf{a}-\textbf{c}) the left ventricular end-diastolic volume (EDV),
	(\textbf{d}-\textbf{f}) the left ventricular end-systolic volume (ESV),
	and (\textbf{g}-\textbf{i}) the left ventricular ejection fraction (EF)
	derived from the nnU-Net segmentation and the manually corrected contours (mcc) are compared with each other in Bland-Altman plots.
	The parameters are compared for cine CMR (\textbf{a}, \textbf{d}, \textbf{g}),
	real-time free-breathing CMR at rest (\textbf{b}, \textbf{e}, \textbf{h}),
	and under exercise stress (\textbf{c}, \textbf{f}, \textbf{i}).
	$P$ values of a paired two-sample t-test are shown in the top left corner of each plot.}
	\label{fig:BA_CF_nnunet}
\end{figure*}

\begin{figure*}[ht]
	\centering
	\includegraphics[width=\textwidth]{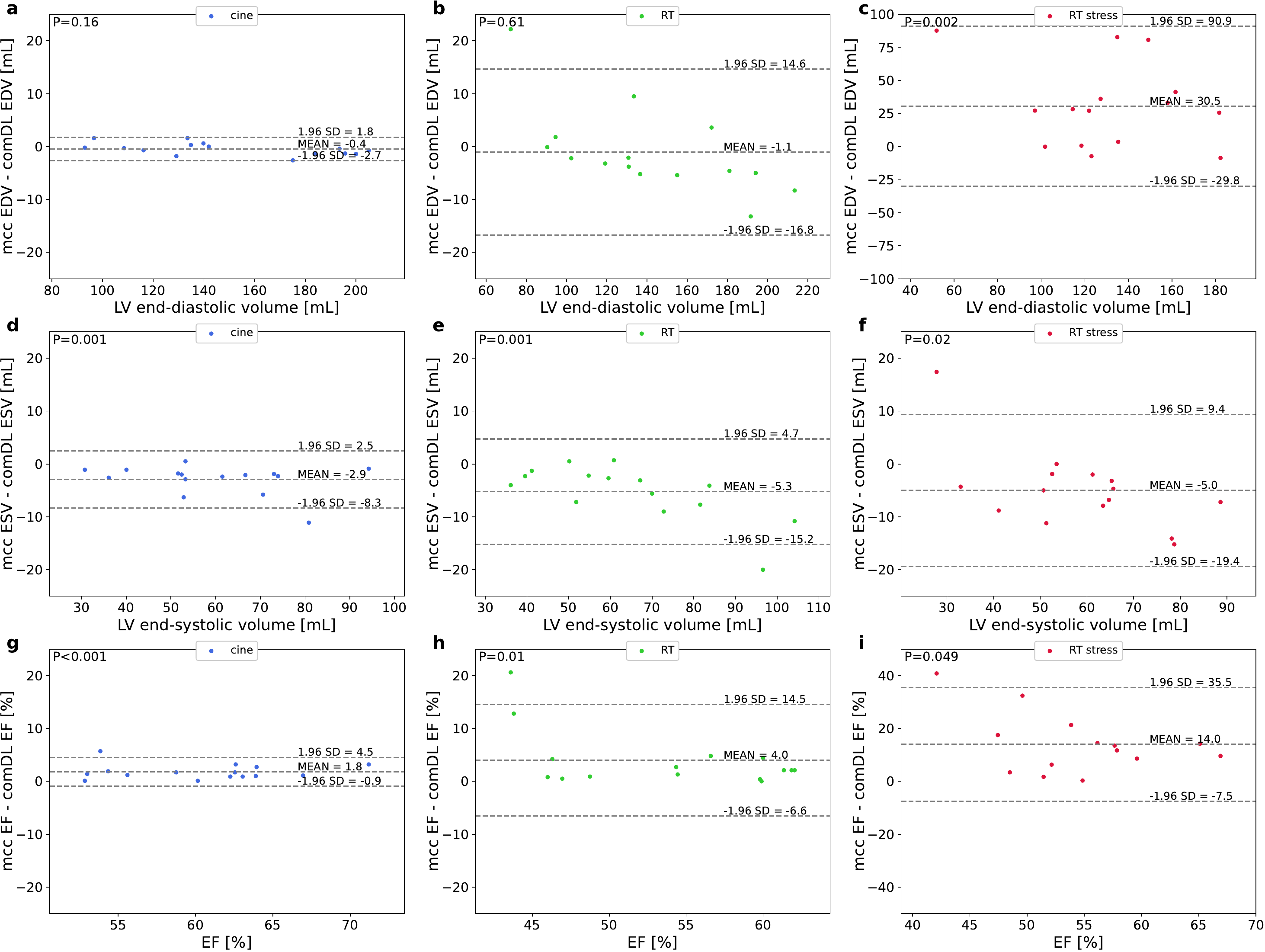}
	\caption{Bland-Altman plots for cardiac function parameters of comDL.
	The cardiac function parameters of (\textbf{a}-\textbf{c}) the left ventricular end-diastolic volume (EDV),
	(\textbf{d}-\textbf{f}) the left ventricular end-systolic volume (ESV),
	and (\textbf{g}-\textbf{i}) the left ventricular ejection fraction (EF)
	derived from the comDL segmentation and the manually corrected contours (mcc) are compared with each other in Bland-Altman plots.
	The parameters are compared for cine CMR (\textbf{a}, \textbf{d}, \textbf{g}),
	real-time free-breathing CMR at rest (\textbf{b}, \textbf{e}, \textbf{h}),
	and under exercise stress (\textbf{c}, \textbf{f}, \textbf{i}).
	One data point for EF at RT stress (\textbf{i}) has been omitted as the EF value derived from comDL contours was negative.
	$P$ values of a paired two-sample t-test are shown in the top left corner of each plot.}
	\label{fig:BA_CF_comDL}
\end{figure*}

\begin{figure*}[ht]
	\centering
	\includegraphics[width=\textwidth]{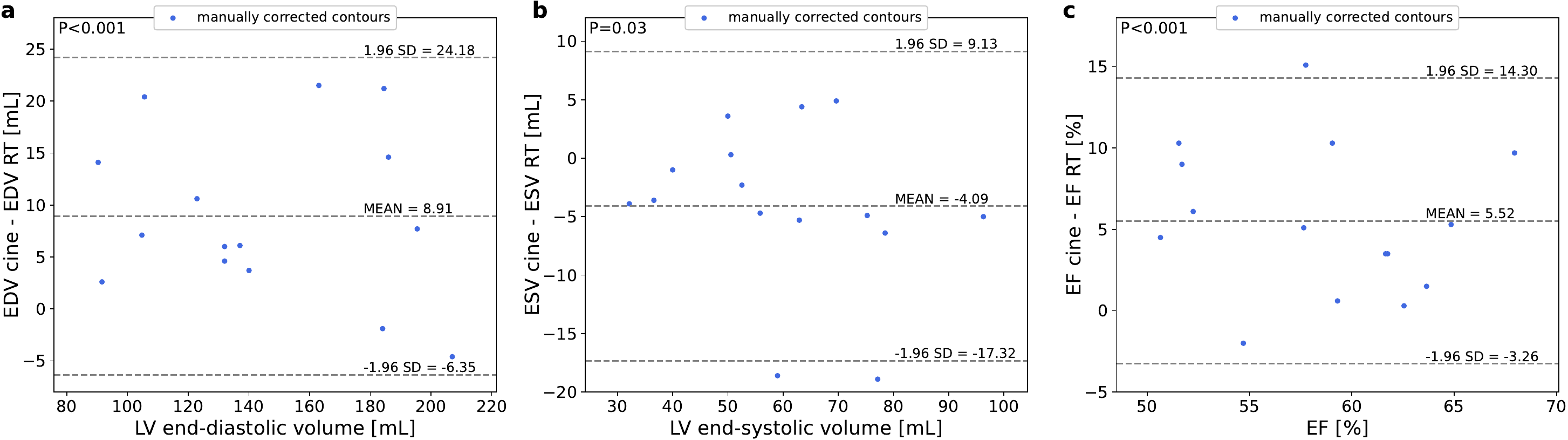}
	\caption{Comparison between cine and real-time CMR cardiac function parameters.
	The cardiac function parameters of (\textbf{a}) the left ventricular end-diastolic volume (EDV),
	(\textbf{b}) the left ventricular end-systolic volume (ESV),
	and (\textbf{c}) the left ventricular ejection fraction (EF)
	derived from manually corrected contours for cine and real-time free-breathing CMR at rest are compared with each other in Bland-Altman plots.
	$P$ values of a paired two-sample t-test are shown in the top left corner of each plot.}
	\label{fig:BA_CF_cine_rt}
\end{figure*}

\begin{figure*}[ht]
	\centering
	\includegraphics[width=\textwidth]{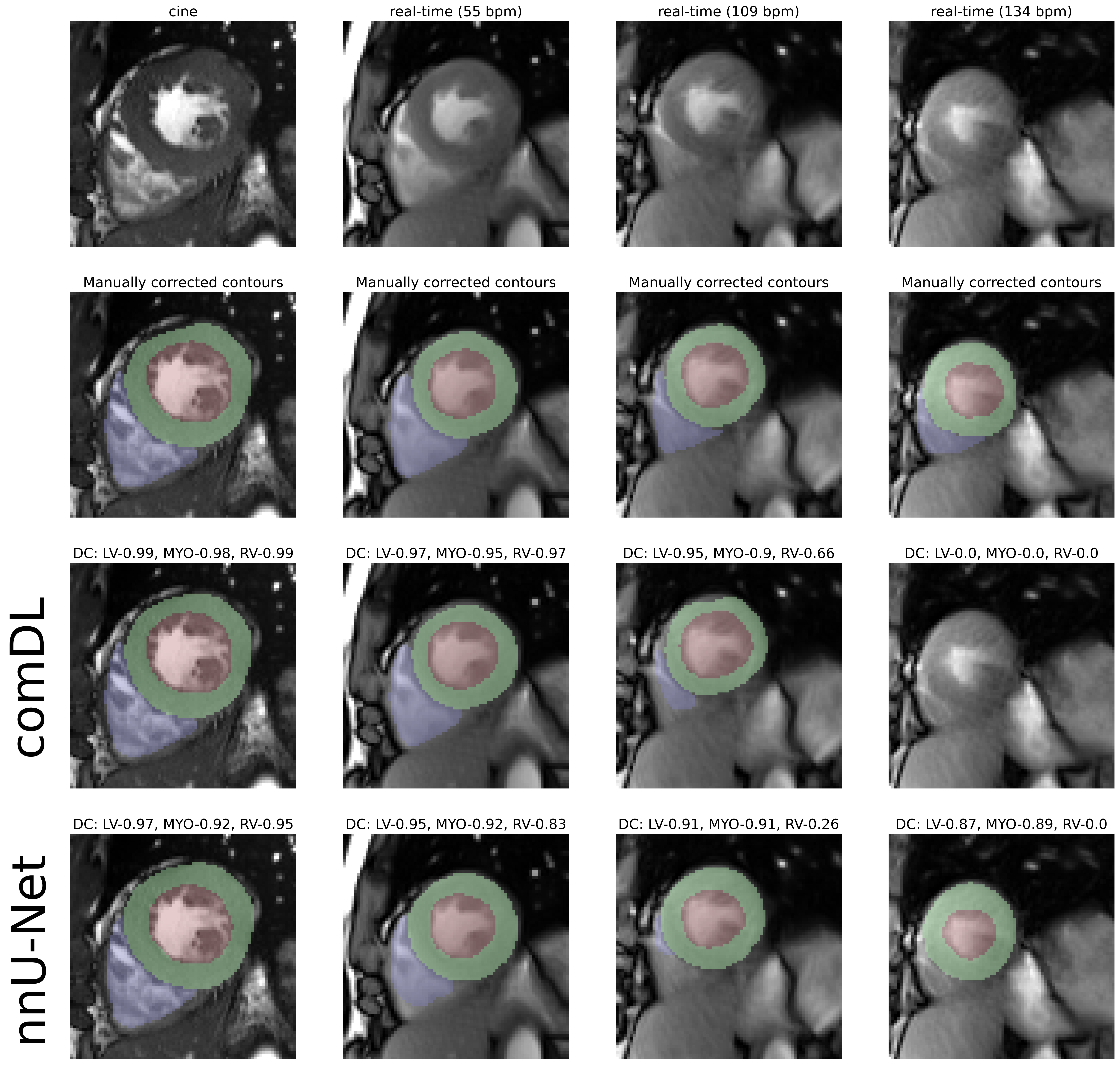}
	\caption{Segmentation failure of right ventricle segmentation outlier of nnU-Net.
	Mid-ventricular slices in ES phase of a volunteer for cine and real-time free-breathing at different heart rates (first row)
	with corresponding manually corrected (second row), comDL (third row), and nnU-Net segmentation (fourth row).
	Accuracy of segmentation is measured with Dice's coefficient (DC). DC for left ventricular endocard (LV), left ventricular myocardium (MYO),
	and right ventricle (RV) are given for each segmentation.
	The images for real-time stress (109 bpm) and real-time maxstress (134 bpm) are segmentation failures of the two outliers of
	RV segmentation of the nnU-Net.
	For these images, the transition between RV and the neighboring liver tissue seems to be too smooth for a successful segmentation.
	The comDL segmentation fails completely for real-time maxstress, while the nnU-Net segments LV and MYO with good accuracy.}
	\label{fig:RV_outlier}
\end{figure*}

\begin{table}[ht]
	\caption[Acquisition parameters for cine and real-time CMR]
	{Acquisition parameters for cine and real-time CMR.
	All volunteers were measured with the same protocol.
	Cine and real-time CMR were acquired using a bSSFP sequence.
	Real-time CMR used iterative image reconstruction with NLINV.}
	\label{tab:acq_param}
	\begin{tabular*}{\textwidth}{@{\extracolsep\fill}lcc}
		\toprule
		& cine	& real-time \\
		\cmidrule(lr){1-3}
		\hline
		Sequence type & Cartesian & radial \\
		Number of images per slice & 25 & 120-150 \\
		Slice thickness [mm] & 6 & 6 \\
		Field of View [mm$^2$] & 340\,x\,276.25 & 256\,x\,256 \\
		Matrix size [pixel] & 256\,x\,208 & 160\,x\,160 \\
		Spatial resolution [mm$^2$] & 1.33\,x\,1.33 & 1.6\,x\,1.6 \\
		Flip angle [degree] & 49 & 23 \\
		Echo time [ms] & 1.53 & 1.28 \\
		Temporal resolution [ms] & 30.1 & 33 \\
		\bottomrule
	\end{tabular*}
\end{table}

\begin{table}[ht]
	\caption[Comparison of 2D nnU-Net predictions on single and stacked images]
	{Comparison of 2D nnU-Net on single and stacked images
	The segmentation accuracy of the 2D nnU-Net inference is compared for the application on single and stacked images.
	Although the 2D nnU-Net is applied on single images, the inference with the 2D nnU-Net features test time augmentation and normalization,
	which take all input data into account.
	To determine the optimal form of application for our task, we applied the 2D nnU-Net on single images and stacks of multiple images
	for cine CMR and real-time free-breathing CMR at rest, under exercise stress, and under maximal exercise stress.
	The table features the mean Dice's coefficients for the left ventricular endocard (LV),
	the left ventricular myocardium (MYO), and the right ventricle (RV) for all volunteers.
	For RT max stress, only data of 12 volunteers were analyzed, because in three cases the image quality was too poor
	to create reasonable reference contours.
	Only images in the end-diastolic and end-systolic phase were segmented.
	In our case, segmentation benefitted from test time augmentation and normalization of single images.
	The differences in segmentation accuracy between the application on single images and stacks of images
	increase for more challenging segmentation tasks.
	For cine CMR, images with the same cardiac phase are stacked along the slice dimension.
	For real-time CMR, images of a time series in a single slice are stacked.}
	\label{tab:nnunet_single_stack}
	\begin{tabular*}{\textwidth}{@{\extracolsep\fill}lcccccc}
		\toprule
		&
		\multicolumn{2}{c}{\bfseries LV} &
		\multicolumn{2}{c}{\bfseries MYO}&
		\multicolumn{2}{c}{\bfseries RV}\\
		n=15& single&stack & single&stack & single&stack \\ \cmidrule(lr){1-7}
		cine 		& 0.952	& 0.951	& 0.907	& 0.907	& 0.904 & 0.898	\\
		RT rest		& 0.943	& 0.912	& 0.885	& 0.859	& 0.896 & 0.849	\\
		RT stress	& 0.921	& 0.833	& 0.850	& 0.775	& 0.826 & 0.674	\\
		RT max stress (n=12)	& 0.909	& 0.788	& 0.826	& 0.748	& 0.788 & 0.598	\\
		\bottomrule
	\end{tabular*}
\end{table}

\begin{table}[ht]
	\caption[Comparison of 2D nnU-Net, 3D nnU-Net, and their ensemble for cine CMR]
	{Comparison of 2D nnU-Net, 3D nnU-Net, and their ensemble
	The segmentation accuracy of the 2D nnU-Net, the 3D nnU-Net, and their ensemble is evaluated for cine CMR of all volunteers.
	Only slices containing the left ventricle have been stacked, similar to the ACDC dataset.
	Only images in the end-diastolic and end-systolic phase were segmented.
	The table features the mean Dice's coefficients for the left ventricular endocard (LV),
	the left ventricular myocardium (MYO), and the right ventricle (RV) for all volunteers.}
	\label{tab:nnunet_ensemble}
	\begin{tabular*}{\textwidth}{@{\extracolsep\fill}lccc}
		\toprule
		n=15 & LV & MYO & RV\\\cmidrule(lr){1-4}
		2D nnU-Net	& 0.951	& 0.907	& 0.898	\\
		3D nnU-Net	& 0.791	& 0.589	& 0.292	\\
		ensemble nnU-Net& 0.926	& 0.852	& 0.833	\\
		\bottomrule
	\end{tabular*}
\end{table}

\begin{table}[ht]
	\caption[Calculated heart rates for real-time free-breathing measurements]
	{Calculated heart rates for real-time measurements.
	Calculated heart rates in beats per minute for real-time free-breathing CMR at rest (RT rest),
	during exercise stress (RT stress), and maximal exercise stress (RT max stress).
	For the calculation of heart rates, we used the three central slices of all slices between the base and apex.
	Heart rates were calculated based on the time span between consecutive end-diastolic phases.
	The mean and standard deviation (in parenthesis) is given for each volunteer.
	Missing entries were not segmented due to poor image quality.}
	\label{tab:vol_heart_rate}
	\begin{tabular*}{\textwidth}{@{\extracolsep\fill}lccc}
		\toprule
		n=15 & RT rest [bpm] & RT stress [bpm] & RT max stress [bpm]\\\cmidrule(lr){1-4}
		vol 01	& 66 (2)	& 113 (4)	& 150 (8)	\\
		vol 02	& 67 (5)	& 111 (7)	& 140 (8)	\\
		vol 03	& 63 (3)	& 114 (4)	& 122 (14)	\\
		vol 04	& 55 (2)	& 109 (7)	& 134 (5)	\\
		vol 05	& 53 (2)	& 117 (6)	& ---		\\
		vol 06	& 68 (1)	& 109 (6)	& 140 (8)	\\
		vol 07	& 73 (3)	& 119 (7)	& 147 (8)	\\
		vol 08	& 77 (3)	& 115 (7)	& 140 (10)	\\
		vol 09	& 75 (3)	& 120 (5)	& 164 (13)	\\
		vol 10	& 68 (3)	& 108 (4)	& 155 (13)	\\
		vol 11	& 65 (4)	& 116 (6)	& ---		\\
		vol 12	& 76 (3)	& 114 (6)	& 162 (11)	\\
		vol 13	& 74 (3)	& 108 (10)	& ---		\\
		vol 14	& 64 (3)	& 111 (6)	& 121 (6)	\\
		vol 15	& 57 (1)	& 107 (6)	& 122 (10)	\\
		\bottomrule
	\end{tabular*}
\end{table}

\begin{table}[ht]
	\caption[Cardiac function parameters for cine and real-time free-breathing CMR measurements]
	{Cardiac function parameters for cine and real-time CMR}
	\label{tab:vol_cardiac_function}
	\begin{tabular*}{\textwidth}{@{\extracolsep\fill}lccc}
		\toprule
		(\textbf{a}) EDV [mL] & & & \\
		n=15 & cine & RT rest & RT stress \\\cmidrule(lr){1-4}
		vol 01	& 128.2	& 117.6	& 119.4	\\
		vol 02	& 141.9	& 138.2	& 145.3	\\
		vol 03	& 92.9	& 90.3	& 95.8	\\
		vol 04	& 193.3	& 178.7	& 176.3	\\
		vol 05	& 204.7	& 209.3	& 189.5	\\
		vol 06	& 135.0	& 129.0	& 137.1	\\
		vol 07	& 140.1	& 134.0	& 135.5	\\
		vol 08	& 173.8	& 152.3	& 178.0	\\
		vol 09	& 115.8	& 95.4	& 118.8	\\
		vol 10	& 97.4	& 83.3	& 108.3	\\
		vol 11	& 199.3	& 191.6	& 194.5	\\
		vol 12	& 108.3	& 101.2	& 101.7	\\
		vol 13	& 183.1	& 185.0	& 182.3	\\
		vol 14	& 134.3	& 129.7	& 128.6	\\
		vol 15	& 195.1	& 173.9	& 174.6	\\
		(\textbf{b}) ESV [mL] & & & \\
		n=15 & cine & RT rest & RT stress \\\cmidrule(lr){1-4}
		vol 01	& 51.8	& 48.2	& 59.5	\\
		vol 02	& 65.6	& 61.2	& 63.7	\\
		vol 03	& 30.2	& 34.1	& 36.5	\\
		vol 04	& 72.1	& 67.2	& 60.2	\\
		vol 05	& 72.8	& 77.7	& 71.1	\\
		vol 06	& 60.3	& 65.6	& 61.3	\\
		vol 07	& 51.4	& 53.7	& 48.2	\\
		vol 08	& 75.3	& 81.7	& 85.0	\\
		vol 09	& 50.7	& 50.4	& 53.5	\\
		vol 10	& 34.8	& 38.4	& 33.1	\\
		vol 11	& 93.8	& 98.8	& 71.1	\\
		vol 12	& 39.5	& 40.5	& 36.7	\\
		vol 13	& 49.7	& 68.3	& 51.6	\\
		vol 14	& 53.5	& 58.2	& 45.7	\\
		vol 15	& 67.7	& 86.6	& 63.3	\\
		(\textbf{c}) EF [\%] & & & \\
		n=15 & cine & RT rest & RT stress \\\cmidrule(lr){1-4}
		vol 01	& 59.6	& 59.0	& 50.2	\\
		vol 02	& 53.7	& 55.7	& 56.2	\\
		vol 03	& 67.5	& 62.2	& 61.9	\\
		vol 04	& 62.7	& 62.4	& 65.8	\\
		vol 05	& 64.4	& 62.9	& 62.5	\\
		vol 06	& 55.3	& 49.2	& 55.3	\\
		vol 07	& 63.4	& 59.9	& 64.4	\\
		vol 08	& 56.7	& 46.4	& 52.3	\\
		vol 09	& 56.2	& 47.2	& 55.0	\\
		vol 10	& 64.2	& 53.9	& 69.4	\\
		vol 11	& 52.9	& 48.4	& 63.4	\\
		vol 12	& 63.5	& 60.0	& 63.9	\\
		vol 13	& 72.8	& 63.1	& 71.7	\\
		vol 14	& 60.2	& 55.1	& 64.5	\\
		vol 15	& 65.3	& 50.2	& 63.7	\\
		\bottomrule
	\end{tabular*}
\end{table}


\begin{thebibliography}{10}

	\bibitem{White_Circ.J._1987}
	H.~D. White, R.~M. Norris, M.~A. Brown, P.~W. Brandt, R.~M. Whitlock, and C.~J.
	  Wild, ``Left ventricular end-systolic volume as the major determinant of
	  survival after recovery from myocardial infarction,'' {\em Circ. J.},
	  vol.~76, no.~1, pp.~44--51, 1987.

	\bibitem{Norris_Eur.HeartJ._1992}
	R.~M. Norris, H.~D. White, D.~B. Cross, C.~J. Wild, and R.~M.~L. Whitlock,
	  ``Prognosis after recovery from myocardial infarction: the relative
	  importance of cardiac dilatation and coronary stenoses,'' {\em Eur. Heart
	  J.}, vol.~13, no.~12, pp.~1611--1618, 1992.

	\bibitem{Uecker_NMRBiomed._2010}
	M.~Uecker, S.~Zhang, D.~Voit, A.~Karaus, K.-D. Merboldt, and J.~Frahm,
	  ``{R}eal-time {MRI} at a resolution of 20 ms,'' {\em NMR Biomed.}, vol.~23,
	  no.~8, pp.~986--994, 2010.

	\bibitem{Feng_Magn.Reson.Med._2013}
	L.~Feng, M.~B. Srichai, R.~P. Lim, A.~Harrison, W.~King, G.~Adluru, E.~V.~R.
	  Dibella, D.~K. Sodickson, R.~Otazo, and D.~Kim, ``{Highly accelerated
	  real-time cardiac cine MRI using k-t SPARSE-SENSE},'' {\em Magn. Reson.
	  Med.}, vol.~70, no.~1, pp.~64--74, 2013.

	\bibitem{Saybasili_Magn.Reson.Imaging_2014}
	H.~Saybasili, D.~A. Herzka, N.~Seiberlich, and M.~A. Griswold, ``Real-time
	  imaging with radial grappa: Implementation on a heterogeneous architecture
	  for low-latency reconstructions,'' {\em Magnetic Resonance Imaging}, vol.~32,
	  no.~6, pp.~747--758, 2014.

	\bibitem{Laubrock_Eur.J.Radiol.Open_2022}
	K.~Laubrock, T.~v. Loesch, M.~Steinmetz, J.~Lotz, J.~Frahm, M.~Uecker, and
	  C.~Unterberg-Buchwald, ``Imaging of arrhythmia: Real-time cardiac magnetic
	  resonance imaging in atrial fibrillation,'' {\em Eur. J. Radiol. Open},
	  vol.~9, 2022.

	\bibitem{Steinmetz_Circ.Cardiovasc.Imaging_2021}
	M.~Steinmetz, T.~Thomas~St\"umpfig, M.~Seehase, A.~Schuster, J.~Kowallick,
	  M.~M\"uller, C.~Unterberg-Buchwald, S.~Kutty, J.~Lotz, M.~Uecker, and P.~T.,
	  ``Impaired exercise tolerance in repaired tetralogy of fallot is associated
	  with impaired biventricular contractile reserve: An exercise-stress real-time
	  cardiovascular magnetic resonance study,'' {\em Circ. Cardiovasc. Imaging},
	  vol.~14, no.~8, 2021.

	\bibitem{Li_Magn.Reson.Imaging_2021}
	Y.~Y. Li, P.~Zhang, S.~Rashid, Y.~J. Cheng, W.~Li, W.~Schapiro, K.~Gliganic,
	  A.-M. Yamashita, M.~Grgas, E.~Haag, and J.~J. Cao, ``Real-time exercise
	  stress cardiac mri with fourier-series reconstruction from golden-angle
	  radial data,'' {\em Magnetic Resonance Imaging}, vol.~75, pp.~89--99, 2021.

	\bibitem{Backhaus_Circ.J._2023}
	S.~J. Backhaus, T.~Lange, E.~F. George, K.~Hellenkamp, R.~J. Gertz, M.~Billing,
	  R.~Wachter, M.~Steinmetz, S.~Kutty, U.~Raaz, J.~Lotz, T.~Friede, M.~Uecker,
	  G.~Hasenfuß, T.~Seidler, and A.~Schuster, ``Exercise stress real-time
	  cardiac magnetic resonance imaging for noninvasive characterization of heart
	  failure with preserved ejection fraction,'' {\em Circ. J.}, vol.~143, no.~15,
	  pp.~1484--1498, 2021.

	\bibitem{Eitel_Curr.Cardiol.Rep._2014}
	C.~Eitel, G.~Hindricks, M.~Grothoff, M.~Gutberlet, and P.~Sommer, ``{C}atheter
	  {A}blation {G}uided by {R}eal-{T}ime {MRI},'' {\em Curr. Cardiol. Rep.},
	  vol.~16, no.~8, pp.~1--7, 2014.

	\bibitem{Unterberg-Buchwald_J.Cardiov.Magn.Reson._2017}
	C.~{Unterberg-Buchwald}, C.~O. Ritter, V.~Reupke, R.~N. Wilke, C.~Stadelmann,
	  M.~Steinmetz, A.~Schuster, G.~Hasenfuß, J.~Lotz, and M.~Uecker, ``Targeted
	  endomyocardial biopsy guided by real-time cardiovascular magnetic
	  resonance,'' {\em J. Cardiov. Magn. Reson.}, vol.~19, no.~1, p.~45, 2017.

	\bibitem{Campbell-Washburn_J.Magn.Reson.Imaging_2017}
	A.~Campbell-Washburn, M.~Tavallaei, M.~Pop, E.~Grant, H.~Chubb, K.~Rhode, and
	  G.~Wright, ``Real-time mri guidance of cardiac interventions,'' {\em J. Magn.
	  Reson. Imaging}, vol.~46, pp.~935--950, 2017.

	\bibitem{Franson_J.Imaging_2021}
	D.~Franson, A.~Dupuis, V.~Gulani, M.~Griswold, and N.~Seiberlich, ``A system
	  for real-time, online mixed-reality visualization of cardiac magnetic
	  resonance images,'' {\em Journal of Imaging}, vol.~7, no.~12, 2021.

	\bibitem{Bernard_IEEETrans.Med.Imag._2018}
	O.~Bernard, A.~Lalande, C.~Zotti, F.~Cervenansky, X.~Yang, P.-A. Heng,
	  I.~Cetin, K.~Lekadir, O.~Camara, M.~A. Gonzalez~Ballester, G.~Sanroma,
	  S.~Napel, S.~Petersen, G.~Tziritas, E.~Grinias, M.~Khened, V.~A. Kollerathu,
	  G.~Krishnamurthi, M.-M. Rohé, X.~Pennec, M.~Sermesant, F.~Isensee,
	  P.~Jäger, K.~H. Maier-Hein, P.~M. Full, I.~Wolf, S.~Engelhardt, C.~F.
	  Baumgartner, L.~M. Koch, J.~M. Wolterink, I.~Išgum, Y.~Jang, Y.~Hong,
	  J.~Patravali, S.~Jain, O.~Humbert, and P.-M. Jodoin, ``Deep learning
	  techniques for automatic mri cardiac multi-structures segmentation and
	  diagnosis: Is the problem solved?,'' {\em IEEE Transactions on Medical
	  Imaging}, vol.~37, no.~11, pp.~2514--2525, 2018.

	\bibitem{Bai_J.Cardiov.Magn.Reson._2018}
	W.~Bai, M.~Sinclair, G.~Tarroni, O.~Oktay, M.~Rajchl, G.~Vaillant, A.~M. Lee,
	  N.~Aung, E.~Lukaschuk, M.~M. Sanghvi, F.~Zemrak, K.~Fung, J.~M. Paiva,
	  V.~Carapella, Y.~J. Kim, H.~Suzuki, B.~Kainz, P.~M. Matthews, S.~E. Petersen,
	  S.~K. Piechnik, S.~Neubauer, B.~Glocker, and D.~Rueckert, ``Automated
	  cardiovascular magnetic resonance image analysis with fully convolutional
	  networks,'' {\em J. Cardiov. Magn. Reson.}, vol.~20, no.~1, p.~65, 2018.

	\bibitem{Bhuva_CircImaging_2019}
	A.~N. Bhuva, W.~Bai, C.~Lau, R.~H. Davies, Y.~Ye, H.~Bulluck, E.~McAlindon,
	  V.~Culotta, P.~P. Swoboda, G.~Captur, T.~A. Treibel, J.~B. Augusto, K.~D.
	  Knott, A.~Seraphim, G.~D. Cole, S.~E. Petersen, N.~C. Edwards, J.~P.
	  Greenwood, C.~Bucciarelli-Ducci, A.~D. Hughes, D.~Rueckert, J.~C. Moon, and
	  C.~H. Manisty, ``A multicenter, scan-rescan, human and machine learning cmr
	  study to test generalizability and precision in imaging biomarker analysis,''
	  {\em Circulation: Cardiovascular Imaging}, vol.~13, no.~3, pp.~684--695,
	  2019.

	\bibitem{Chen_Front.Cardiovasc.Med._2020}
	C.~Chen, C.~Qin, H.~Qiu, G.~Tarroni, J.~Duan, W.~Bai, and D.~Rueckert, ``Deep
	  learning for cardiac image segmentation: A review,'' {\em Front. Cardiovasc.
	  Med.}, vol.~7, no.~25, 2020.

	\bibitem{Shoaib_ComputationalIntelligenceandNeuroscience_2023}
	M.~A. Shoaib, J.~H. Chuah, R.~Ali, K.~Hasikin, A.~Khalil, Y.~C. Hum, Y.~K. Tee,
	  S.~Dhanalakshmi, and K.~W. Lai, ``An overview of deep learning methods for
	  left ventricle segmentation,'' {\em Computational Intelligence and
	  Neuroscience}, vol.~2023, 2023.

	\bibitem{Morales_J.Cardiov.Magn.Reson._2023}
	M.~A. Morales, S.~Yoon, A.~Fahmy, F.~Ghanbari, S.~Nakamori, J.~Rodriguez,
	  J.~Yue, J.~A. Street, D.~A. Herzka, W.~J. Manning, and R.~Nezafat, ``Highly
	  accelerated free-breathing real-time myocardial tagging for exercise
	  cardiovascular magnetic resonance,'' {\em J. Cardiov. Magn. Reson.}, vol.~25,
	  no.~1, p.~56, 2023.

	\bibitem{Yang_BioMedRes.Int._2019}
	F.~Yang, Y.~Zhang, P.~Lei, L.~Wang, Y.~Miao, H.~Xie, and Z.~Zeng, ``A deep
	  learning segmentation approach in free-breathing real-time cardiac magnetic
	  resonance imaging,'' {\em BioMed Res. Int.}, 2019.

	\bibitem{Qi_Physiol.Meas._2022}
	Y.~Qi, F.~Wang, J.~J. Cao, and Y.~Y. Li, ``A deep learning approach to
	  real-time volumetric measurements without image reconstruction for
	  cardiovascular magnetic resonance,'' {\em Physiol. Meas.}, vol.~43, no.~10,
	  2022.

	\bibitem{Uecker_Magn.Reson.Med._2010}
	M.~Uecker, S.~Zhang, and J.~Frahm, ``{N}onlinear inverse reconstruction for
	  real-time {MRI} of the human heart using undersampled radial {FLASH},'' {\em
	  Magn. Reson. Med.}, vol.~63, no.~6, p.~1456–1462, 2010.

	\bibitem{Zhang_J.Cardiov.Magn.Reson._2010}
	S.~Zhang, M.~Uecker, D.~Voit, K.~Merboldt, and J.~Frahm, ``{R}eal-time
	  cardiovascular magnetic resonance at high temporal resolution: radial {FLASH}
	  with nonlinear inverse reconstruction,'' {\em J. Cardiov. Magn. Reson.},
	  vol.~12, no.~1, p.~39, 2010.

	\bibitem{Isensee_Nat.Methods_2021}
	F.~Isensee, P.~F. J\"ager, S.~A.~A. Kohl, J.~Petersen, and K.~H. Maier-Hein,
	  ``nnu-net: a self-configuring method for deep learning-based biomedical image
	  segmentation,'' {\em Nat. Methods}, vol.~18, pp.~203--211, 2021.

	\bibitem{Isensee_git_2023}
	F.~Isensee, ``Welcome to the new nnu-net!.''
	  \url{https://github.com/MIC-DKFZ/nnUNet/tree/master}, 2023.
	\newblock Accessed: 2023-10-26.

	\bibitem{Greenberg_J.Thorac.Imag._2000}
	S.~Greenberg, ``Assessment of cardiac function: Magnetic resonance and computed
	  tomography,'' {\em J. Thorac. Imag.}, vol.~15, no.~4, pp.~243--251, 2000.

	\bibitem{Mahnken_Rofo-Fortschr.Rontg._2004}
	A.~Mahnken, R.~G\"unther, and G.~Krombach, ``The basics of left ventricular
	  functional analysis with mri and msct,'' {\em Rofo-Fortschr. Rontg.},
	  vol.~176, no.~10, pp.~1365--1379, 2004.

	\bibitem{Amirrajab_Lect.Notes.Comput.Sc._2022}
	S.~Amirrajab, Y.~Al~Khalil, J.~Pluim, M.~Breeuwer, and C.~M. Scannell,
	  ``Cardiac mr image segmentation and quality control in the presence of
	  respiratory motion artifacts using simulated data,'' in {\em Statistical
	  Atlases and Computational Models of the Heart. Regular and CMRxMotion
	  Challenge Papers} (O.~Camara, E.~Puyol-Ant{\'o}n, C.~Qin, M.~Sermesant,
	  A.~Suinesiaputra, S.~Wang, and A.~Young, eds.), (Cham), pp.~466--475,
	  Springer Nature Switzerland, 2022.

	\bibitem{Ronneberger_Lect.NotesComput.Sc._2015}
	O.~Ronneberger, P.~Fischer, and T.~Brox, ``U-net: Convolutional networks for
	  biomedical image segmentation,'' in {\em Medical Image Computing and
	  Computer-Assisted Intervention -- MICCAI 2015} (N.~Navab, J.~Hornegger, W.~M.
	  Wells, and A.~F. Frangi, eds.), (Cham), pp.~234--241, Springer International
	  Publishing, 2015.

	\bibitem{El-Taraboulsi_ArtificialIntelligenceintheLifeSciences_2023}
	J.~El-Taraboulsi, C.~P. Cabrera, C.~Roney, and N.~Aung, ``Deep neural network
	  architectures for cardiac image segmentation,'' {\em Artificial Intelligence
	  in the Life Sciences}, vol.~4, p.~100083, 2023.

	\bibitem{Zhou_IEEETrans.ImageProcessing_2023}
	H.-Y. Zhou, J.~Guo, Y.~Zhang, X.~Han, L.~Yu, L.~Wang, and Y.~Yu, ``nnformer:
	  Volumetric medical image segmentation via a 3d transformer,'' {\em IEEE
	  Trans. Image Processing}, vol.~32, pp.~4036--4045, 2023.

	\bibitem{Hosny_Nat.Rev.Cancer_2018}
	A.~Hosny, C.~Parmar, J.~Quackenbush, L.~H. Schwartz, and H.~J. W.~L. Aerts,
	  ``Artificial intelligence in radiology,'' {\em Nat. Rev. Cancer}, vol.~18,
	  pp.~500--510, 2018.

	\bibitem{ESR_InsightsImaging_2019}
	E.~S. of~Radiology~(ESR), ``What the radiologist should know about artificial
	  intelligence - an esr white paper,'' {\em Insights Imaging}, vol.~10, no.~44,
	  pp.~500--510, 2019.

	\bibitem{Reardon_Nature_2019}
	R.~S., ``Rise of robot radiologists,'' {\em Nature}, vol.~576, pp.~54--58,
	  2019.

	\bibitem{Mongan_Radiol.Artif.Intell._2020}
	J.~Mongan, L.~Moy, and C.~E. Kahn, ``Checklist for artificial intelligence in
	  medical imaging (claim): A guide for authors and reviewers,'' {\em Radiology:
	  Artificial Intelligence}, vol.~2, no.~2, p.~e200029, 2020.

	\bibitem{Alabed_Front.Cardiovasc.Med._2022}
	S.~Alabed, A.~Maiter, M.~Salehi, A.~Mahmood, S.~Daniel, S.~Jenkins, M.~Goodlad,
	  M.~Sharkey, M.~Mamalakis, V.~Rakocevic, K.~Dwivedi, H.~Assadi, J.~M. Wild,
	  H.~Lu, D.~P. O'Regan, R.~J. van~der Geest, P.~Garg, and A.~J. Swift,
	  ``Quality of reporting in ai cardiac mri segmentation studies - a systematic
	  review and recommendations for future studies,'' {\em Front. Cardiovasc.
	  Med.}, vol.~9, 2022.

	\bibitem{Maiter_Front.Radiol._2023}
	A.~Maiter, M.~Salehi, A.~J. Swift, and S.~Alabed, ``How should studies using ai
	  be reported? lessons from a systematic review in cardiac mri,'' {\em Front.
	  Radiol.}, vol.~3, 2023.

	\bibitem{Sander_Proc.SPIE_2019}
	J.~Sander, B.~D. de~Vos, J.~M. Wolterink, and I.~Išgum, ``Towards increased
	  trustworthiness of deep learning segmentation methods on cardiac mri,'' in
	  {\em Medical Imaging 2019: Image Processing} (E.~D. Angelini and B.~A.
	  Landman, eds.), vol.~10949, (San Diego, California, United States),
	  p.~1094919, SPIE, 2019.

\end{thebibliography}
\end{document}